# High sensitivity differential Magneto-Optical Imaging with a compact Faraday-modulator

Pabitra Mandal[1], Debanjan Chowdhury[1,2], S. S. Banerjee[1,a], T. Tamegai[3]

[1]Department of Physics, Indian Institute of Technology, Kanpur-208016, India

[2]Present address: Department of Physics, Harvard University, Cambridge Massachusetts 02138, USA

[3] Department of Applied Physics, The University of Tokyo, Hongo, Bunkyo-ku, Tokyo 113-8656, Japan

**Abstract:**

We present here the design of a sensitive Compact Faraday-modulator (CFM) based optical magnetometer for imaging the distribution of weak local magnetic fields inside hysteretic magnetic materials. The system developed has a root mean square (rms) noise level of 50 mG.Hz$^{-1/2}$ at a full frame rate of 1 frame per second with each frame being of size 512 × 512 pixels. By measuring the local magnetic field distribution in different superconducting samples we show that our magnetometer provides an order of magnitude improvement in the signal to noise (*S/N*) ratio at low fields as compared to ordinary magneto-optical imaging technique. Moreover, it provides the required sensitivity for imaging the weak magnetization response near a superconducting transition where a number of other imaging techniques are practically unviable. An advantage of our CFM design is that it can be scaled in size to fit into situations with tight space constraints.

## I. INTRODUCTION

Magneto - Optical Imaging (MOI)[1] is a useful technique to map out local magnetic field (*B*) distribution of a magnetic material. In an MOI setup, a magneto-optically active film having a high Verdet's constant, *V*, is placed on the surface of the sample. The magneto-optically active layer is an in-plane magnetized Bismuth substituted Yttrium Iron Garnet film grown on (100) oriented Gadolinium Gallium Garnet substrates via isothermal liquid phase epitaxy[2]. When linearly polarized light passes through the magneto-optically active film, it undergoes Faraday rotation by an angle $\theta_F$ which is proportional to the magnetic field experienced by the film[3,4,5]. By capturing the intensity of the Faraday rotated light, information of the magnetic field experienced by the film is obtained. For performing magneto-optical imaging in the reflection mode a thin reflecting Al layer (~ 50 nm thick) is deposited on the magneto-optically active film. The reflecting layer is placed in contact with the sample during MO imaging. It is this reflecting Al layer which is responsible for doubling the optical path length traversed by light in the magneto-optically active medium.

---

[a] email: satyajit@iitk.ac.in

Depending on the magnetic properties of the sample, there is a spatial variation of the local magnetic field $B_z(x,y)$ across the sample surface ($z$ is defined to be the direction perpendicular to the plane of the sample) which in turn generates a non-uniform distribution of Faraday rotated light intensity ($I$) across the sample. Using a charge coupled device (CCD) camera we image the Faraday rotated light intensity distribution ($I(x,y)$) across the magneto-optical (MO) image. Using this MO image we determine the local magnetic field distribution across the sample. Within the linear response regime of the magneto-optically active film[2], the intensity of the (reflected) Faraday rotated linearly polarized light rotated through an angle $\theta_F = B_z V (2d)$, is given by[1,2,3] $I = I_0 \sin^2(2VB_z d)$, where $d$ is the thickness of the magneto-optically active layer (the factor of 2 is due to doubling of the optical path length). For typical values of the parameters $V \sim 10^{-3}$ °μm$^{-1}$ Oe$^{-1}$, $d = 5$ μm and for low $B_z$ (in the range of few 100 G), the Faraday rotated light intensity is well approximated by: $I(x,y) \propto (B_z(x,y))^2$. Thus, by measuring $I(x,y)$ one can determine $B_z(x,y)$ with reasonable accuracy. However, for very low values of $B_z(x,y)$, the corresponding $I(x,y)$ is buried in the background noise. Conventional MOI suffers from a variety of sources of noise like photon shot noise, dark current noise of the CCD camera, electronic noise in the CCD camera, intensity fluctuations of the light source used in the MOI setup etc. All these different sources contribute significantly to the noise in the MO images captured, thereby reducing the sensitivity with which regions with weak local magnetic fields present inside the sample can be detected with a conventional MOI setup. In conventional MOI setups, typical average local magnetic fields of the order of 5 - 10 G can be detected.

In recent times, differential magneto-optical (DMO) imaging technique has achieved high field sensitivity, viz., the DMO technique detects *changes* in local magnetic field of the order of 10 mG in a background field of few hundred Gauss. The DMO technique detects with high sensitivity small changes in $B_z(x,y)$, viz., $\delta B_z(x,y)$ produced in response to small changes in externally applied magnetic field or temperature. In this technique, an average differential image is constructed by taking the difference between images captured by slightly modulating either the temperature ($T$)[6,7,8,9] or the applied magnetic field ($H$)[7,9,10]. An average of $n$ magneto-optical images ($I$) is captured at $H+\delta H$ or at $T+\delta T$ ($\delta H \sim 1$ Oe or $\delta T \sim 0.2$ K), $\langle I(H+\delta H \text{ or } T+\delta T) \rangle_n$ where $\langle .... \rangle_n$ denotes the average over $n$ number of MO images, and one also determines the mean of $n$ images captured at $H$ or $T$, $\langle I(H \text{ or } T) \rangle_n$. Using these, one constructs $\langle \delta I_n \rangle_m$ where $\delta I_n = \langle I(H+\delta H \text{ or } T+\delta T) \rangle_n - \langle I(H \text{ or } T) \rangle_n$ and $\langle .... \rangle_m$ indicates the average of the differential image $\delta I_n$ computed over $m$ iterations. It turns out[6,7] that $\langle \delta I_n \rangle_m \propto \delta B_z(x,y)$. The high sensitivity DMO technique has been employed to image physical phenomena associated with small changes in local magnetic field, $\delta B_z(x,y)$, occurring in the background of a finite local magnetic field distribution, i.e. where $B_z(x,y) \gg \delta B_z(x,y)$. For instance, this technique was used for imaging the phenomenon of first-order vortex-lattice melting[6,7,9,10] in a type-II superconductor. Due to vortex-lattice melting, jumps in the local density of vortices equivalent to a small jump in $\delta B_z \sim 100$ mG riding on a large background of $B_z \sim 200$ G were detected using the DMO imaging technique. In recent times a magneto-optical imaging setup with high local magnetic field resolution has successfully imaged individual vortices[11] in a superconductor. While the conventional DMO technique offers high sensitivity for detecting differential changes ($\delta B_z(x,y)$) in the local magnetic field, this procedure inherently involves capturing differential images by modulating the external field or temperature. Therefore, it is applicable for materials in those $H$, $T$ ranges where the magnetization response is reversible. To avoid the above limitation, we present here the design of a setup employing differential imaging technique using a compact Faraday modulator (CFM) wherein the principle of modulating the polarization of the incident light[12] is used instead of modulating the external $H$ or $T$ to obtain differential images. Our differential technique using the CFM detects with high sensitivity the local magnetic field distribution, $B_z(x,y)$ in a sample rather than detecting $\delta B_z(x,y)$ in conventional DMO technique. Our setup has an rms (root mean square) noise level of 50 mG.Hz$^{-1/2}$ per pixel at a minimum frame rate of 1 frame per second, at a resolution of 512 × 512 pixels. Using our set up we compare measurements of $B_z(x,y)$ within the hysteretic regime of superconductors with those obtained using conventional MOI, to show at least an order of magnitude improvement in the signal to noise ratio ($S/N$). We also show that our CFM based differential imaging technique holds the potential to be used as a sensitive optical magnetometer to measure the local as well as bulk magnetization hysteresis loops of hysteretic magnetic materials. Additionally we are also able to image and resolve weak magnetization signals especially close to $T_c$ of a superconductor where a number of other imaging techniques become insensitive.

## II. DESIGN OF THE COMPACT FARADAY MODULATOR (CFM)

A schematic diagram of our magneto-optical imaging setup incorporating the CFM is shown in fig. 1(a). The imaging set up is similar to a conventional MOI setup, consisting of a 100 W incoherent light source (L) (for performing imaging under noisy background conditions we have intentionally employed an ordinary halogen lamp as the light source), a commercial Carl-Zeiss Axiotech vario polarized light microscope with low strain and long working distance Epiplan-Neofluar objective (O) with 10x magnification. Imaging is done at a wavelength of 550 nm where the value of $V$ is maximum for the magneto-optically active garnet film. The incident light passes through a green filter (550 nm ± 10 nm) and then through a film linear polarizer (P). The polarizer forms a part of the CFM shown in fig.1(a). Shown in fig. 1(b) is an actual image of the CFM (built out of a transparent acrylic sheet which has dimensions 7.5 × 4.5 × 0.9 cm$^3$, thickness of the sheet is 0.9 cm). A cylindrical bore is drilled through the acrylic sheet along its thickness with a diameter 3.71 cm. The cylindrical bore has a solenoid C made with copper wire wound on a spool made also out of the transparent acrylic sheet. The solenoid has an open bore with inner diameter of 1.81 cm and an outer diameter 3.68 cm. Inside the open bore of the solenoid C, a combination of a circular piece of a film linear polarizer (P) of diameter 1.81 cm is placed followed by a transparent Faraday active film (F) (viz., a Faraday active film *devoid of* the reflecting Al layer to enable operation in the *transmission mode*).

The rectangular plane of the CFM with dimensions 7.5 cm × 4.5 cm is placed perpendicular to the incident light beam, the thickness (0.9 cm) of the CFM is along the incident light such that the beam of green light emerging out of the filter is linearly polarized by the polarizer (P) before passing through the transparent Faraday active film (F). A current ($I_F$) through the coil (C) produces a uniform magnetic field across the bore of the solenoid in the CFM in which F is placed. The magnetic field experienced by F in the CFM causes the transiting linearly polarized light to undergo a Faraday rotation upon passing through F. Depending on the strength of the applied magnetic field experienced by F in the CFM, a desired amount of rotation of the plane of linear polarization ($\theta_p$) can be produced. Note that, $\theta_p$ is controlled by the strength of the magnetic field (and hence, by $I_F$ sent through C) experienced by F. The Faraday active film (F) used in the CFM has in-plane magnetization which rotates out of plane with $B_z$ within the solenoid (C). These magneto-optically active films have low coercive field and possess large magnetic domains, due to which these Faraday active films are employed in magneto-optical imaging[2]. The physical dimensions of the film (F) used in the CFM are smaller than the typical size of the magnetic domains in these films at room temperature.

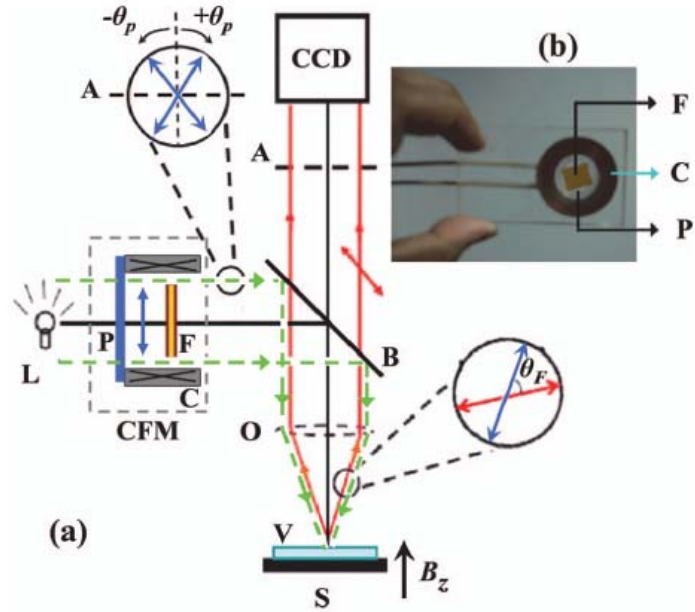

**Figure 1. (a) Schematic diagram of the CFM combined with conventional MOI setup. L, ordinary incoherent light source; P, film polarizer; F, transparent Faraday active film; C, solenoid coil; O, objective lens; B, beam splitter; A, analyzer; V, Faraday active indicator film with reflecting Al layer and S, sample. The double headed arrows indicate the plane of polarization of the beam. The blow up on the incident ray (drawn by blue arrows) illustrates the modulation of the polarization angle $\theta_p$ of the incident light ray while passing through the CFM. The blow up on the combination of the incident and reflected ray (drawn by red arrow) illustrates the rotation of polarization $\theta_F$ upon reflection from the Faraday active indicator placed on the sample.**
**(b) A picture of the CFM: F, Faraday active film; C, coil and P, film polarizer.**

As a result, there is almost no domain activity found within the field of view of our MO imaging setup with the CFM. Additionally, due to the low coercive field of F, the presence of any large in-plane component of magnetic field acting on F inside the solenoid (C) of the CFM would result in the nucleation of in-plane magnetized magnetic domains. To avoid nucleating these domains we ensure that the plane of F is aligned perpendicular to the magnetic field inside the CFM. Near the inner walls of C, as the magnetic field is not completely perpendicular to the F plane, in-plane magnetic domains are generated near the edges of F. However these magnetic domains do not affect the imaging since the incident light is confined to illuminate the central region of F which is devoid of domains. In our CFM, we have avoided using Faraday active films with physical damage or defects since these regions act as favorable sites for nucleating magnetic domains. Furthermore in our modulator design, F is maintained at a nearly constant temperature and strain free environment which ensures that we do not have any domain activity in F during the course of its operation. The compact CFM in principle has the potential to be scaled down in size to conveniently fit into constrained tight spaces in the optical path of an imaging setup. In our setup, the CFM fits inside an empty additional thin rectangular slot of dimensions 7.5 cm × 4.5 cm × 0.9 cm, provided in the commercial Carl Zeiss microscope used for our experiment. The compactness of our CFM design as compared to earlier attempts[12] was achieved due to the use of a sensitive (5 μm thick) Faraday active film (F) with a large Verdets constant ($V = 8.5 \times 10^{-4}$ deg. $\mu m^{-1} Oe^{-1}$, at 300K). Earlier polarimeters[12] are up to 15 cm long compared to just 0.9 cm in our case. Additional compactness of design was achieved by optimizing the wire gauge in the solenoid and number of turns so as to enable an appreciable rotation of the polarized light passing through the CFM while keeping the size of the coil in the CFM compatible within the space constraints in our imaging setup.

The optimum configuration of the solenoid was attained with a copper wire of 0.3 mm diameter having 400 turns. Using a Hall probe, we have measured a field of 180 Oe at the center of the solenoid in the CFM, when a current ($I_F$) of 1 Amp is sent through the coil. The field homogeneity is about 1% along the inner diameter of the solenoid and $10^{-3}$ % variation over the thickness (~ 5 μm) of the Faraday active film placed inside the solenoid of the CFM. By sending ± $I_F$ current through the coil, the polarized light emerging out of CFM is modulated by ±$\theta_p$ (see zoomed in portion of the schematic in fig.1(a)). The light beam from the CFM before reaching the beam splitter (B in fig.1) passes through an aperture (not shown in fig.1). The light beam emerging from the CFM has an almost Gaussian like intensity profile with a full width at half maximum of about 2 cm. The aperture of diameter 3 mm which is centered about the location of the peak intensity in the Gaussian beam, transforms the incident light into one with an almost uniform intensity profile across its cross section. Furthermore the optical components in our system are aligned to ensure the absence of any gradients in light intensity illuminating the region being imaged. To confirm this, fig.2 shows a single image of a uniformly illuminated Faraday active film placed in zero externally applied magnetic field at room temperature, captured on our setup with the CFM (with $\theta_p = 0$ deg). We show in fig.2 two plots of the intensity variations measured along the two lines (along the *x* and *y* directions) indicated in the image. Clearly, over a distance of about 800 μm x 800 μm, we do not observe any significant curvature in the mean intensity profile or any intensity gradients in the profiles. This indicates an almost uniform illumination over the imaging field of view and negligible influences arising from any non-uniformity in the light beam intensity distribution.

The Faraday modulated light (green dashed line in fig.1 schematic), after emerging from CFM continues onto a beam splitter (B) where it is partially reflected and is focused onto the sample via the objective (O). On top of the sample (S), another Faraday active indicator film (V) with a reflecting Al layer is placed. The Faraday rotated light reflected from the sample passes through (B) and then encounters the analyzer (A) placed at right angle to the pass direction of the film polarizer (P) in the CFM through which the incident light beam is transmitted. The intensity of the Faraday rotated light is captured via the CCD camera. Using a Peltier cooled (-100ºC) 16 bit CCD camera (Andor iXon) with 512×512 pixels (with each pixel of size 16 × 16 μm$^2$), images are captured. All parameters in our setup, viz., frame rate, angle of polarization ($\theta_p$), exposure time of the CCD camera and number of images to average, number of pixels on the CCD (out of the 512 × 512 pixels on the CCD array) constituting an image, can be programmatically interfaced and controlled. Note that for the measurements reported in this paper with our CFM, the sample (S) from which the light is reflected is placed in an evacuated chamber in which the temperature can be varied from 10 K to 350 K. The sample (S) is placed in a magnetic field $H$ applied using a water cooled copper wire wound solenoid coil from which a maximum $H$ = 500 Oe is attainable.

### III. PRINCIPLE FOR OBTAINING DIFFERENTIAL IMAGES WITH CFM

As discussed in the previous section, the CFM modulates the polarization of the incident light beam, viz., it rotates the polarization of the incident beam of light reaching the sample being investigated by $\pm\theta_p$ (see the zoomed in section showing the state of polarization of the incident beam in fig. 1(a)). In our differential technique employing the CFM, images are captured synchronously by alternately rotating the polarization of the incoming linearly polarized light with a magnetic field applied on the Faraday active film (F) inside the CFM (cf. fig. 1), viz., we obtain $\langle \delta I_n \rangle_m = \langle \langle I_+ \rangle_n - \langle I_- \rangle_n \rangle_m$ where $\langle I_+ \rangle_n$ and $\langle I_- \rangle_n$ represent n – image average when the polarization of the incident linearly polarized light is modulated by an angle by $+\theta_p$ and $-\theta_p$ respectively. If we choose the analyzer pass axis to be along the horizontal direction, the incident polarization is modulated and measured w.r.t. the crossed position to the analyzer pass axis, i.e., vertical direction (see the zoomed in section in fig. 1(a)).

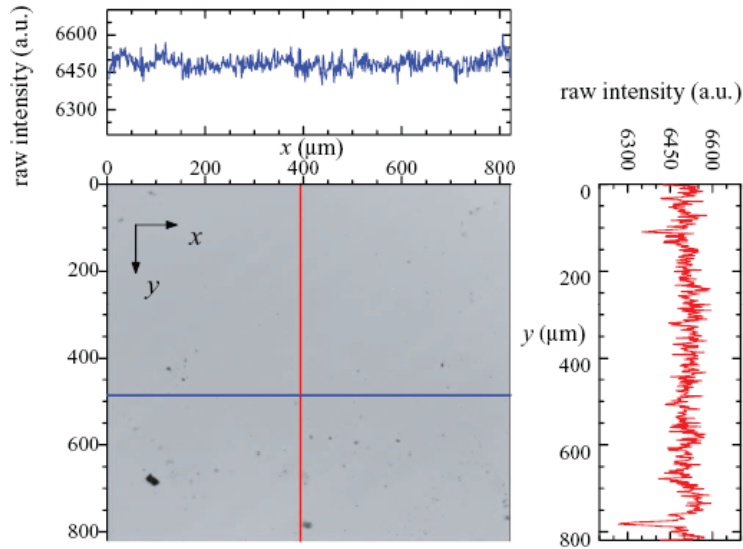

**Figure 2. Image of the Faraday active indicator placed on top of the sample captured via CCD camera at room temperature. The graphs are showing two raw intensity profiles across two orthogonal directions shown by red and blue lines on the image. Uniform illumination is observed in the intensity profiles. Some gray and black spots in the image indicate defects on the indicator.**

Thus, if $\theta$ is the angle of polarization of the incident beam, the intensity after passing through the analyzer is given by: $I = I_{off} + I_0 \sin^2(VB_z 2d + \theta) \approx I_{off} + I_0(VB_z 2d + \theta)^2$, for small angles, where $I_0$ is the incident intensity. The first term in the equation, $I_{off}$, is the background illumination where the major contributions to background illumination are associated with the extinction ratio (~$10^{-4}$) of the crossed linear polarizer in our setup and the dark current associated with the CCD chip. For $\theta = +\theta_p$, a series ($n$) of images are captured producing an averaged magneto-optical (MO) image $\langle I_+(x,y) \rangle_n \approx \langle I_{off}(x,y) \rangle_n + \langle I_0(x,y) \rangle_n (VB_z(x,y)2d + \theta_p)^2$. For $\theta = -\theta_p$, the same procedure is followed to generate $\langle I_-(x,y) \rangle_n \approx \langle I_{off}(x,y) \rangle_n + \langle I_0(x,y) \rangle_n (VB_z(x,y)2d - \theta_p)^2$. Note that our technique is applicable for close to equilibrium conditions of magnetization in the sample, i.e., when there is negligible dynamics of magnetization. Therefore in the above expressions, $B_z(x,y)$ denotes a time independent value of the local magnetic field at a given location ($x,y$) in a sample when all other ambient conditions like, temperature, applied magnetic field, etc., remain unchanged.

The main sources of noise which limit the sensitivity for detecting small $B_z(x,y)$ in our set up are the temporal fluctuations in Faraday rotated light intensity. This noise is associated with photon shot noise, the electronic noise of the CCD camera, the intensity fluctuations of the light source and vibrations in the setup. We find that by averaging over a sufficiently large number ($n$) of images (typically ~ 20 image frames with an exposure time set to ~ 1 second for each frame), most temporal fluctuations are averaged out and $\langle I_{off}(x,y) \rangle_n$, $\langle I_0(x,y) \rangle_n$ do not exhibit any temporal drift. From $\langle I_+ \rangle_n$ and $\langle I_- \rangle_n$ we construct the differential image, $\delta I_n = \langle I_+ \rangle_n - \langle I_- \rangle_n$, and the above procedure is repeated $m$ times to obtain the averaged differential image

$$\langle \delta I_n \rangle_m = \langle \langle I_+ \rangle_n - \langle I_- \rangle_n \rangle_m = kB_z(x,y) + \Delta \qquad (1)$$

where $k = 8I_0(x,y)Vd\theta_p$ is a calibration constant for our setup and $\Delta$ is a term associated with fluctuations in the differential intensity. As discussed in the context of fig.2 above, since the $I_0(x,y)$ is almost uniform over the imaged area (typical imaged area doesn't exceed 500 μm x 500 μm), therefore we use $k$ to be constant over this area. Averaging over a large number of images (by choosing $n$ and $m$ to be large), the random fluctuations in the mean differential intensity $\langle \delta I_n \rangle_m$ at each pixel ($x,y$) can be significantly reduced to a value such that $\Delta \ll kB_z(x,y)$, thereby offering a large signal to noise ratio. As the differential procedure is time consuming for low light conditions (like those for low $B_z(x,y)$), therefore in our set up we measure a time averaged $B_z(x,y)$ distribution across a sample. LabView is used to programmatically interface and control the CCD camera as well as the CFM to yield $\langle \delta I_n \rangle_m$. According to Eq. 1, the differential intensity is proportional to $B_z$, implying that the present differential technique directly provides information about the magnitude as well as polarity of $B_z(x,y)$. Furthermore, in the presence of a noisy background ($\Delta$), the $\langle \delta I_n \rangle_m$ offers higher sensitivity for detecting a very small $B_z(x,y)$ due to the linear dependence of $\langle \delta I_n \rangle_m$ on $B_z(x,y)$, in comparison to a conventional MOI technique[2] where the mean intensity is related to $k'B_z^2(x,y)$, $k'$ being a constant. To determine the calibration factor $k$ in equation (1), we measure the MO intensity ($I^{out}$) averaged over an area of 10×10 μm$^2$ far away from the sample where the sample's magnetization response has a negligible effect on the local $B_z(x,y)$, so that one can use $B_z(x,y) \approx H$, the externally applied field. The proportionality constant $k$ is easily determined from a linear fit of $I^{out}$ vs. $H$ plot constructed from a set of images captured at different $H$ as discussed in the calibration section. The advantage of the present technique is that one obtains an improved signal to noise ($S/N$) ratio for determining the $B_z(x,y)$ by modulating the polarization of the incoming light rather than measuring the differential increment in $B_z(x,y)$, i.e., $\delta B_z(x,y)$ with a modulation of the $H$ or $T$ applied to the sample as in conventional DMO. Unlike the conventional DMO technique, the present technique is suited for studying materials with irreversible magnetization response without disturbing the thermomagnetic history of the sample during the measurement process.

## IV. OPTIMIZATION
### A. Optimizing the operating current ($I_F$) in the CFM

To optimize the operating current ($I_F$) through the CFM, the $S/N$ ratio is calculated systematically for different $I_F$. A single crystal of high $T_c$ superconductor $Bi_2Sr_2CaCu_2O_8$ (BSCCO)[13] of dimensions (0.8 × 0.5 × 0.03 mm$^3$) with superconducting transition temperature $T_c$ = 90 K was used for this characterization. we measure the nature of the local magnetic field distribution in the interior of the superconductor using our CFM based MOI set up for low applied fields. The superconductor is maintained at a field $H$ = 36 Oe (|| c – crystallographic axis of the single crystal) and at a temperature $T$ = 30 K (zero field cooling (ZFC)). The MO images of the BSCCO sample are shown in figs.3(c) and (d). In these figures, the region inside the superconductor where $B_z$ is smaller than that outside the sample (due to the diamagnetic response of the superconductor), appears with darker contrast compared to the exterior of the sample with a finite non zero $B$ (= 36 G). The light intensity at a pixel in the 16 - bit CCD camera is measured on a gray scale with shades ranging between values 0 (black) and $2^{16}$ (white). . The noise ($N$) is calculated from the standard deviation in the raw intensity counts averaged over an 80×80 μm$^2$ area shown as a box in fig. 3(d), outside the sample in the image. The red graph in the inset of fig. 3(a) shows the raw intensity, $\langle \delta I \rangle$, profile ($\propto B_z$) calculated from an MO image obtained by our MO setup incorporating the CFM, along a line drawn across the sample shown in fig.3(d). Inside the sample, $\langle \delta I \rangle$ drops to a low value due to the diamagnetic shielding response of the superconductor. This graph is obtained with $2mn$ = 10 no. of averaging ($m$ = 1, n = 5, where $m$, $n$ are already defined in section III, equation 1) and $I_F$ = 0.4A. The blue curve is a smoothened curve through the fluctuating data, indicating the average behavior of $\langle \delta I \rangle$ across the superconducting sample. The dome shaped nature of this $\langle \delta I \rangle$ distribution inside the superconductor is expected from a dome shaped $B_z$ distribution due to screening currents circulating on the sample edges and negligible bulk screening current[14]. Note that at the boundaries of the sample, the enhanced intensity is associated with the edge screening currents. The signal ($S$) is the height of the dome as shown in fig. 3(a) inset. Some of the large fluctuations in the actual data in fig. 3(a) inset (see at the arrow locations) are due to physical defects, like scratches on the Faraday active film (V) placed on the top of the sample.

The main panel of fig. 3(a) shows the behavior of the $S/N$ vs. $I_F$, constructed out of MO images captured with $m$ = $n$ = 10 (i.e., 200 image averaging ( = $2mn$), cf. equation 1). We note that the $S/N$ ratio increases with $I_F$, improving by as much as 60% with increasing $I_F$ until an optimum value of 0.5 A ( = 90 Oe field in the CFM), beyond which the $S/N$ ratio saturates and does not change significantly with $I_F$. We usually work with an $I_F$ = 0.4 A to minimize heating in the solenoid coil C in the CFM as heating induces thermal strain in the Faraday active film giving rise to strain induced birefringence effects. The lower $I_F$ also prevents damaging the solenoid coil in the CFM as there is no cooling mechanism. A current of 0.4 A in the CFM produces a rotation by $\theta_p = 0.3°$ of the plane of polarization of the light beam being transmitted through it.

### B. Optimizing the number of averaging

The improvement in the *S/N* ratio with total number of averaging (= $2mn$) is investigated and optimized with $I_F = 0.4$ A. Figure 3(b) shows that the *S/N* ratio improves with the number of averaging. From the standard deviation of the pixel values in the box region (5 pixel × 5 pixel) shown outside the sample in fig.3(d), we determine the rms noise level, while the signal is determined by the method discussed in section IV(A). From the graph in fig. 3(b), we observe that initially the *S/N* ratio improves sharply with the no. of averaging till $2mn = 100$, beyond which the improvement is gradual. We note that there is an improvement in the *S/N* ratio by almost one order of magnitude at $2mn = 300$ compared to that at $2mn = 10$.

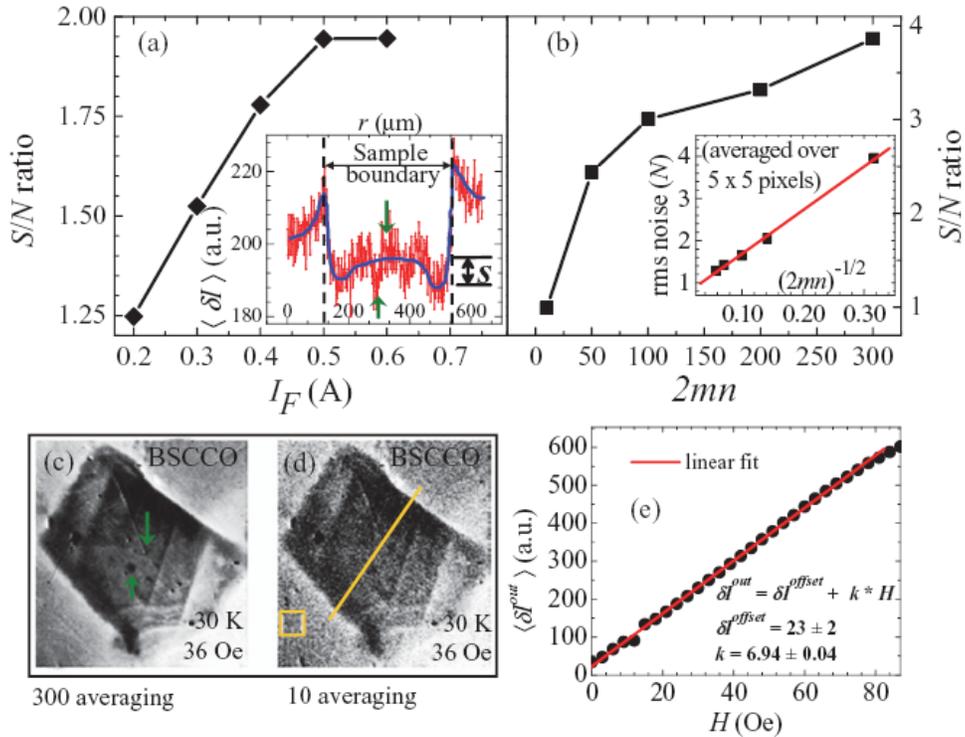

**Figure 3.** (a) Variation of *S/N* ratio with $I_F$ through the CFM. Inset shows the intensity profile $\langle \delta I \rangle$ obtained from the CFM based MOI across a line draw through the Bi$_2$Sr$_2$CaCu$_2$O$_8$ (BSCCO) superconductor shown in fig.3(d) (cf. text for details). (b) Variation of *S/N* ratio with no. of averaging ($2mn$). The inset shows the linear behavior of noise (*N*) vs. $2mn^{-1/2}$ curve. (c) and (d) shows MO images obtained with $2mn = 300$ and 10 respectively (cf. text for details) of BSCCO sample with dimensions (0.8 × 0.5 × 0.03 mm$^3$), measured at 30 K (ZFC) with a field of 36 Oe applied parallel to the c – crystallographic axis of the single crystal. (e) Calibration curve: Black filled circles are experimental data points of mean intensity (over a small area, see text) $\langle \delta I^{out} \rangle$ far outside the sample vs. *H*. The red curve is the linear fit to extract the calibration factor.

The effect of enhanced *S/N* ratio is clearly visible from a comparison of fig. 3(c) and fig. 3(d), where MO image of BSCCO at 30 K with $2mn = 300$ (fig.3(c)) can be compared with MO image obtained with $2mn = 10$ (fig.3(d)). The image with $2mn = 10$ in fig. 3(d) looks grainy and noisier while the sharpness and contrast in the image has significantly improved with more averaging (i.e., $2mn = 300$) in fig.3 (c). Inset of fig. 3(b) shows a linear increase in the rms noise with $1/\sqrt{2mn}$ for our CFM, implying that the noise level decreases linearly with the square root of the no. of averaging ($2mn$) as expected for photon shot noise. We note that the value of the intercept per pixel which is the rms noise per pixel for a large no. of averaging is about $\sim 1/25$ for our set up.

## V. Calibration and Results
### A. Calibration

The calibration curve for the CFM is shown by filled circles in fig. 3(e), obtained from a set of images captured with different $H$ for $I_F = 0.4$ A, $2mn = 200$. This curve shows the variation of the differential intensity, $\langle \delta I^{out}_n \rangle_m$ with an applied magnetic field ($H$) determined in an area of $100 \times 100$ μm$^2$ region on the Faraday active indicator film far away from the sample boundary. The linear behavior of $\langle \delta I^{out}_n \rangle_m$ as a function of $H$ is as expected from eqn. 1 in section III, where far away from the sample, $B_z \sim H$. The slope gives the calibration factor $k = 6.94 \pm 0.04$ (Gray scale).G$^{-1}$, which is easily obtained by a least square fitting of the data (shown by the straight-line through the data points).

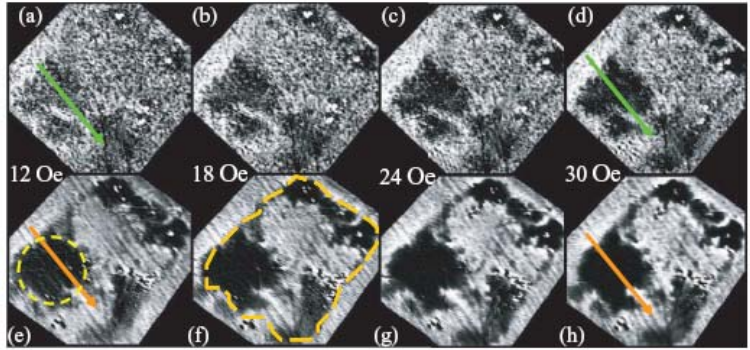

**Figure 4. Comparison between ordinary and CFM based MOI technique: Upper panel ((a)-(d)) shows MO images with ordinary MOI at different $H$ (12 Oe, 18 Oe, 24 Oe and 30 Oe, as indicated in the images), obtained without using the CFM. The lower panel ((e)-(h)) shows images obtained with CFM at same $H$ using same no. of averaging, $2mn = 200$. The measurement is performed at 13 K with $H \parallel$ c – crystallographic axis of the single crystals of CaFe$_{1.94}$Co$_{0.06}$As$_2$ superconductor (with dimensions $0.66 \times 0.47 \times 0.064$ mm$^3$) having $T_c = 17$ K. The region encircled by yellow dashed circle in fig. 4(e) identifies a Meissner region on the sample. The dashed line in fig. 4(f) identifies the boundary of the sample in the MO image.**

The fit follows a form $\langle \delta I^{out}_n \rangle_m = \langle \delta I^{offset}_n \rangle_m + kH$, where the term $\langle \delta I^{offset}_n \rangle_m$ arises from a slight error in setting the analyzer (A, cf. fig.1(a)) in the extinction position w.r.t. the pass axis of the polarizer P. The $\langle \delta I^{offset}_n \rangle_m$ is subtracted from the measured MO intensity with the CFM prior to the calibration of an image to determine $B_z(x,y)$. Using the calibration factor, we estimate from the inset of fig.3(b) that with $2mn = 100$ images and the typical CCD camera exposure time set to about 1 second, the rms noise per pixel for our system is $\sim 50$

mGauss.Hz$^{-1/2}$, with a full frame (512 × 512 pixels) rate of 1 fps (frames per second) at low applied magnetic fields. Compared to the earlier design[12], the rms noise floor level of our CFM has improved by almost two orders of magnitude, while the full frame (512 × 512 pixels) rate for capturing images is down by one order of magnitude. The use of a slower CCD camera causes the lowering of the frame rate for our setup. For the present CCD camera, it is worthwhile to mention that a comparable (or even faster) frame rate as compared to the earlier design[12] can be obtained while maintaining a similar rms noise per pixel by selecting a subset of the 512 × 512 pixels on the CCD array to compose a single image. In such instances a subset p × q pixels (with p & q < 512) centered on the region of interest in the image formed out of the 512 × 512 pixels on the CCD chip are selected via the control driver of the CCD camera. In this way the p × q pixels frame is captured at a faster rate by the CCD camera as compared to a frame composed of 512 × 512 pixels while maintaining a comparable rms noise level. Note that in this procedure the read out and electronic noise associated with the CCD camera remains almost unchanged. We achieve a maximum frame rate of about 100 fps for 10 × 10 pixels selected on the CCD camera.

### B. Results: Detecting the Meissner response of a superconductor at low applied magnetic field and at temperatures close to the superconducting to normal state transition.

To demonstrate the improvement in the *S/N* ratio over the conventional MOI technique[2], we investigate the MO images in a new class of superconductor, i.e., in the single crystals of CaFe$_{1.94}$Co$_{0.06}$As$_2$ superconductor[15] (with dimensions 0.66×0.47×0.064 mm$^3$, $T_c$ (0) ~ 17K). The sample was zero field cooled down to 13 K and the MO images were captured at different applied *H* (∥ c). Figure 4 shows some representative images at 12 Oe, 18 Oe, 24 Oe and 30 Oe with conventional MOI technique (upper panel, by averaging each image 200 times) and with the CFM (lower panel) with 2*mn* = 200. For this study we choose low applied magnetic fields and higher temperature of 13 K which is close to $T_c$(~17K) of the superconductor. We intentionally chose a high temperature (close to $T_c$) and low field regime as the magnetic response of the superconductor is expected to be low due to weak gradients in the local magnetic flux distribution inside it. Due to the weak magnetic response, it becomes harder to discern features in the local magnetic field distribution with conventional MOI. The dark regions in fig. 4(f) in the images correspond to strongly diamagnetic, Meissner shielded regions in the superconductor with *B* = 0 G. The sample boundary is shown by broken (yellow) lines in the same image. At the outset, observe that at 12 Oe while the superconducting sample is hardly visible with ordinary MOI technique (c.f. fig. 4(a)), the sample is clearly visible at the same *H* and *T* in fig. 4(e) which is captured by employing the CFM. The improvement in contrast in the images in the lower panel as compared to the upper panel suggests a significant improvement in the *S/N* ratio for the MO images captured with the CFM. The regions inside the sample where the gray scale contrast is almost the same as that outside the sample are the regions where the magnetic flux has penetrated almost uniformly inside the sample. Note that the grayish lines running diagonally across the images (both across the sample as well as outside the sample) in the lower panel are associated with linear defects on the Faraday active indicator film placed on the sample. The defects on the indicator film are also clearly discernable using the CFM based differential technique (lower panel of fig.4) while they are not visible in the conventional MO images (upper panel of fig.4).

The improvement in the *S/N* ratio for an image captured using the CFM as compared to conventional MOI technique is illustrated further by comparing the $B_z(r)$ profiles in fig. 5(a) and 5(b). The $B_z(r)$ profiles are determined along the line passing through the Meissner region of the superconductor at 12 Oe in figs. 4(a) and 4(e). The vertical dashed lines in fig.5 represent the approximate boundary of the Meissner region encountered along the lines drawn in the images in figs. 4(a) and 4(e). From the $B_z(r)$ plot in fig. 5(a) which is associated with conventional MOI at 12 Oe, it is difficult to identify the location of the Meissner region boundary where one expects to see a sharp drop in $B_z$ to nearly zero compared to a region outside the sample where $B_z$ is finite (12 G).

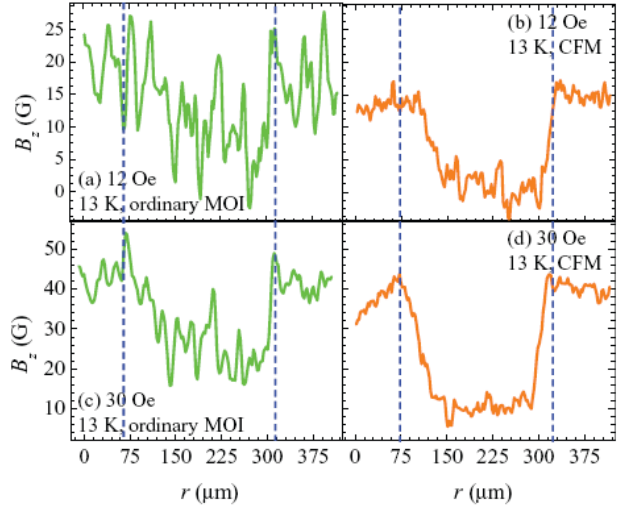

Figure 5. Comparison of $B_z$ ($r$: distance) profiles calculated from the MO images obtained via ordinary MOI and CFM based MOI. The line along which the profiles are obtained are shown in fig. 4 (cf. text for detail): $B_z$ profile at $H$ = 12 Oe (a) via ordinary MOI technique and (b) using CFM based MOI. (c) $B_z$ profile at $H$=30 Oe by ordinary technique and (d) by CFM at same $H$. The vertical dashed lines indicate the sample boundaries.

In the same figure, note that the expected $B_z(r)$ profile is buried within noise level ($N \sim 15$ G) and S/N ratio in this image is 1.3, viz., $N \approx S$. On the other hand, upon using the differential technique with the CFM shown in fig. 5(b), the noise level drops significantly resulting in an improved S/N ratio of about 12.7 and correspondingly the drop in $B_z$ upon crossing the boundary of the Meissner region from outside of the superconductor is clearly discernable for the measurements performed under identical conditions. Obviously, due to an improved S/N ratio, it has become possible to detect the Meissner response visually in the image of the sample with the CFM (cf. fig. 4(e)) whereas with ordinary MOI technique (cf. fig. 4(a)) it is almost not detectable at low fields ($H$ = 12 Oe). A similar behavior in the $B_z(r)$ profile is also observed for higher fields. For example, at $H$ = 30 Oe we can clearly detect a sharp drop in $B_z$ at the two boundaries of the sample in fig. 5(d) with the CFM, but with the ordinary MOI, see fig. 5(c), it is hardly observable. Here, one obtains a S/N ratio of 23.1 with the CFM which is again one order of magnitude higher than that in the ordinary MOI technique where S/N ratio is only 2.5, signifying the noise level to be comparable with the signal in ordinary MOI. Finally, from the slope of the $B_z(r)$ profile near the sample edges at 30 Oe using the CFM technique (cf. fig. 5(d)), we have calculated the superconducting screening current density, $J_c$ $\sim dB_z/dr \sim 5\times10^3$ A/cm$^2$, which is comparable to a superconducting critical current density $J_c$ of the order of $10^4$ A/cm$^2$ reported in the literature[16] for these class of superconductors. Some peculiarities of the local magnetization response in CaFe$_{1.94}$Co$_{0.06}$As$_2$ have been recently discussed in ref.[15].

We have also attempted to image the magnetization response very close to the superconducting transition temperature ($T_c$) of this $CaFe_{1.94}Co_{0.06}As_2$ superconductor. Usually, near the $T_c$ of a superconductor, it becomes difficult to detect the magnetization response as the shielding response becomes very weak. In fig. 6(a), at 16K which is just 1K below $T_c \sim$ 17K of the superconductor, a weak diamagnetic magnetization response from a small portion of the material is captured with CFM technique with a very small applied field $H = 3$ Oe.

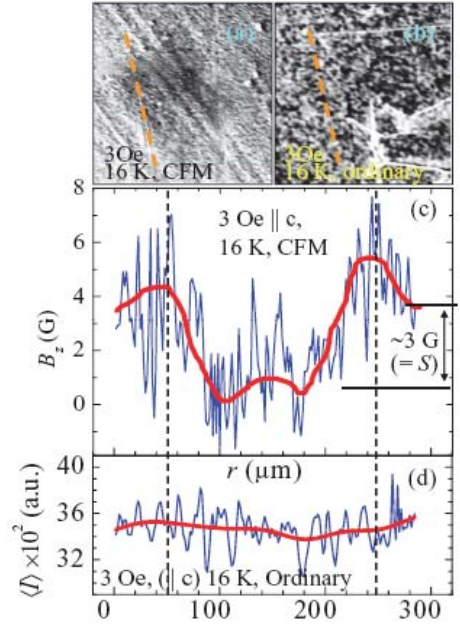

**Figure 6. Detection of Meissner response at 16K (very close to the $T_c$ (= 17K)), $H$ = 3 Oe with both the CFM and ordinary MOI technique: MO images captured (a) via CFM technique and (b) with ordinary MOI (cf. text for detail). (c) The blue curve shows the $B_z(r)$ profile calculated for CFM technique along a line shown in fig. (a). The red thick curve is a guide to eye showing the mean behavior of the actual $B_z(r)$ profile (blue curve). (d) Raw intensity profile for the ordinary MO image (cf. fig. 6(b)) along the same line. The thick red curve is the mean behavior of the actual intensity profile.**

The image in fig. 6(a) is captured with the CFM with $2mn = 100$. The $B_z(r)$ profile across a line (shown in fig.6(a)) in the image captured using the CFM technique is shown in fig. 6(c). The blue curve shows the actual data and the red curve is a guide to the eye showing the mean behavior of the actual data. The dashed vertical lines indicate the sample boundary. Despite the significant amount of noise in the field profile, there is an unmistakable evidence of relatively large screening currents on the sample edges where $B_z$ is large with the characteristic of dome shaped profile[14] inside the sample, which is reminiscent of a similar feature in fig. 3(a) inset. In fig. 6(c) we obtain an S/N ratio ~ 2.35, implying that the signal is higher than the noise. On the other hand, the field profile in fig. 6(d) determined across the line in the conventional MO image in fig.6(b) where the line is drawn at the same location as in fig. 6(a) shows that the superconducting response of the sample is completely buried in the background noise. The thick red curve in fig.6 (d) shows that the mean behavior is almost flat and featureless unlike fig. 6(c). Therefore, our CFM based MOI technique is useful at low fields and high *T*'s close to $T_c$ of the superconductor where detection of weak magnetization signal with most other techniques is practically unviable.

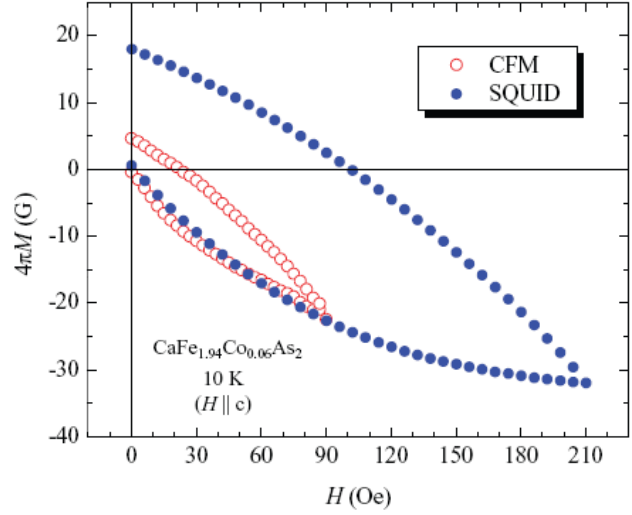

**Figure 7.** Comparison at 10K between the bulk $4\pi M$ vs. $H$ loops obtained via conventional SQUID magnetometer (filled circles) and the loop calculated for the whole sample area out of a set of MO images captured via CFM (cf. text for details).

Finally, we use our CFM based magnetometer to calculate the bulk magnetization hysteresis response of a superconducting $CaFe_{1.94}Co_{0.06}As_2$ single crystal. At 10 K, we captured MO images using the CFM magnetometer at different $H$ (|| c) starting from zero to 90 Oe and returning back to zero. Thereafter we calculate the two quadrant $\langle B_z - H \rangle_A$ vs. $H$ hysteresis loop with the $B_z(x,y)$ determined from the MO images using [17]

$$\langle B_z - H \rangle_A = \frac{\int_A [B_z(x,y) - H] dxdy}{\int_A dxdy},$$ where $A$ is integration area which is the full sample area

(viz., the region within the yellow dashed boundary marked in fig.4(f)). We also measured the bulk magnetization response of the same sample using the Quantum Design, Superconducting Quantum Interference Device (SQUID) based magnetometer. We find that to match the $\langle B_z - H \rangle_A$ data obtained using the CFM with the $4\pi M$ value measured with the SQUID magnetometer, we need to incorporate only an empirical geometrical factor $g$ of the order of one, viz., we compare $g\langle B_z - H \rangle_A$ with the SQUID magnetization data. Figure 7 shows in blue solid circles the bulk $4\pi M$ vs. $H$ curve for the sample at 10 K measured with the SQUID magnetometer. Similar bulk loops have been reported for Pnictides using bulk magnetization measurement techniques like the SQUID magnetometer[16]. On the same figure we add (open red circles) $g\langle B_z - H \rangle_A$ vs. $H$ (where we have used a $g = 0.65$ to match the CFM data with the SQUID data) obtained using our CFM based optical magnetometer. Figure 7 suggests a close match with the magnetization behavior determined using a commercially available SQUID magnetometer with that obtained from our CFM based optical magnetometer. Our CFM based optical magnetometer offers the advantage of sensitively measuring the magnetization response locally and also quantitatively mapping out (with micron scale spatial resolution) the field distribution across a material. Moreover it allows us to study the evolution of this field distribution as a function of applied field and temperature under situations where ordinary MOI has limited sensitivity. While we have used superconducting samples for our study, our system may be employed in the study of any hysteretic magnetic material where imaging under low light condition is required.

**VI. CONCLUSION**

In conclusion, we have developed an optical magnetometer using a compact (only 9 mm thick) Faraday modulator (CFM) which is capable of sensitively measuring as well as imaging weak local magnetic field distributed across a sample. The high sensitivity offered by the CFM based magnetometer enables us to capture MO signal under situations where the sensitivity of the ordinary MOI technique is limited. The differential technique employed in the CFM based magnetometer provides at least a ten fold improvement in the signal to noise ratio in addition to information of the field polarity, as compared to ordinary MOI technique. The CFM based MOI system designed by us also offers easy control on the exposure time, frame rate, polarization angle in the CFM and the number of images to average, independently.

**ACKNOWLEDGEMENTS**

S. S. Banerjee thanks CSIR; DST; IIT K - INDIA for funding support. S. S. Banerjee thanks Prof. S. K. Dhar and Prof. A. Thamizhavel, TIFR, Mumbai, India for the Pnictide single crystals.

References:

[1] *Magneto-Optical Imaging*, NATO Science Series II, Vol. 142, edited by T. H. Johansen and D. V. Shantsev (Kluwer Academic, The Netherlands, 2004).

[2] L. E. Helseth, R. W. Hansen, E. I. Il'yashenko, M. Baziljevich, and T. H. Johansen, Phys. Rev. B **64**, 174406 (2001); L. E. Helseth, A. G. Solovyev, R. W. Hansen, E. I. Il'yashenko, M. Baziljevich, and T. H. Johansen, *ibid.* **66**, 064405 (2002).

[3] Polyanskii A A, Feldmann D M and Larbalestier D C 2003 Magneto-Optical Characterization Techniques *Handbook of Superconducting Materials* Vol. 2, ed D A Cardwell and D S Ginley (Bristol, UK: Institute of Physics, Publishing UK) pp. 1551-1567.

[4] Jooss Ch, Albrecht J, Kuhn H, Leonhardt S and Kronmuller H, 2002 *Rep. Prog. Phys.* **65**, 651-788.

[5] Jaivardhan Sinha, Shyam Mohan, S. S. Banerjee, Subhendu Kahaly and G. Ravindra Kumar, Phys. Rev. E **77**, 046118 (2008).

[6] A. Soibel, Eli Zeldov, Michael Rappaport, Yuri Myasoedov, Tsuyoshi Tamegai, Shuuichi Ooi, Marcin Konczykowski and Vadim B. Geshkenbein, Nature (London) **406**, 282 (2000).

[7] S. S. Banerjee, A. Soibel, Y. Myasoedov, M. Rappaport, E. Zeldov, M. Menghini, Y. Fasano, F. de la Cruz, C. J. van der Beek, M. Konczykowski, and T. Tamegai, Phys. Rev. Lett. **90**, 087004 (2003).

[8] M. Menghini, Yanina Fasano, F. de la Cruz, S. S. Banerjee, Y. Myasoedov, E. Zeldov, C. J. van der Beek, M. Konczykowski, and T. Tamegai, Phys. Rev. Lett. **90**, 147001 (2003).

[9] S. S. Banerje, S. Goldberg, A. Soibel, Y. Myasoedov, M. Rappaport, E. Zeldov, F. de la Cruz, C. J. van der Beek, M. Konczykowski, T. Tamegai, and V.M. Vinokur, Phys. Rev. Lett. **93**, 097002 (2004).

[10] M. Yasugaki, K. Itaka, M. Tokunaga, N. Kameda, and T. Tamegai, Phys. Rev. B **65**, 212502 (2002).

[11] P. E. Goa, Harald Hauglin, Michael Baziljevich, Eugene Il'yashenko, Peter L Gammel and Tom H Johansen, Supercond. Sci. Technol. **14**, 729 (2001); P. E. Goa, H. Hauglin, Å. A. F. Olsen, M. Baziljevich, and T. H. Johansen, Rev, Sci. Instrum., **74,** 141 (2003).

[12] Rinke J. Wijngaarden, K. Heeck, M. Welling, R. Limburg, M. Pannetier, K. van Zetten, V. L. G. Roorda, and A. R. Voorwinden, Rev. Sci. Instrum. **72**, 2661 (2001).

[13] S. Ooi, T. Shibauchi and T. Tamegai, *Physica C* **302**, 339 (1998).

[14] Zeldov E, Larkin A I, Geshkenbein V B, Konczykowski M, Majer D, Khaykovich B, Vinokur V M and Shtrikman H, *Phys. Rev. Lett.* **73** 1428 (1994).

[15] 15P. Mandal, G. Shaw, S. S. Banerjee, N. Kumar, S. K. Dhar, and A. Thamizhavel, Europhys. Lett. **100**, 47002 (2012).

[16] R. Prozorov, M. A. Tanatar, B. Roy, N. Ni, S. L. Bud'ko, P. C. Canfield, J. Hua, U. Welp, and W. K. Kwok, Phys. Rev. B **81**, 094509 (2010).

[17] E. Bartolome, X. Granados, A. Palau, T. Puig, X. Obradors, C. Navau, E. Pardo, A. Sánchez, H. Claus, Phys. Rev. B 72, 024523 (2005); D. Majer, E. Zeldov, and M. Konczykowski, Phys. Rev. Lett. 75, 1166 (1995).